\documentclass[showpacs, pra,twocolumn,preprintnumbers ,amsmath, amssymb, superscriptaddress, aps]{revtex4-2}
\usepackage{color}
\usepackage{amsmath,amssymb}
\usepackage{pifont}
\usepackage{amssymb}  
\usepackage{bbold}
\usepackage{float}
\usepackage{subfloat}
\usepackage[caption=false]{subfig}
\usepackage{tikz}
\usepackage{makecell}
\usepackage{subfig}
\usepackage{pifont}   
\usepackage{graphicx} 
\graphicspath{{Figures/}}
\usepackage{dcolumn}  
\usepackage{bm}       
\usepackage{multirow} 
\usepackage{placeins}
\usepackage[colorlinks]{hyperref}
\usepackage{mathtools}
\usepackage{appendix}

\newcommand{\bb} {\color{blue}}

\def \be{\begin{align}}
	\def \ee{\end{align}}
\def \bea{\begin{eqnarray}}
	\def \eea{\end{eqnarray}}

\begin{document}
	\title{Field-tunable spin–valley transport in monolayer MoS$_2$}
	
\author{Kamal Azaidaoui}
\affiliation{Laboratory of Theoretical Physics, Faculty of Sciences, Choua\"ib Doukkali University, PO Box 20, 24000 El Jadida, Morocco}
\author{Hocine Bahlouli}
\affiliation{Physics Department and IRC Advanced Materials$,$
	King Fahd University
	of Petroleum $\&$ Minerals$,$
	Dhahran 31261$,$ Saudi Arabia}
\author{Clarence Cortes}
\affiliation{Vicerrector\'ia de Investigaci\'on y Postgrado, Universidad de La Serena, La Serena 1700000, Chile}  
\author{David Laroze}
\affiliation{Instituto de Alta Investigación, Universidad de Tarapacá, Casilla 7D, Arica, Chile}
\author{Ahmed Jellal}
\email{a.jellal@ucd.ac.ma}
\affiliation{Laboratory of Theoretical Physics, Faculty of Sciences, Choua\"ib Doukkali University, PO Box 20, 24000 El Jadida, Morocco}

\begin{abstract}
	We study field-controlled spin-valley transport in monolayer MoS$_2$ through a single electrostatic barrier and a uniform off-resonant elliptically polarized irradiation. Starting from the massive Dirac Hamiltonian with intrinsic spin-orbit coupling, we use a high-frequency Floquet expansion to obtain an effective static model with a laser-renormalized mass (gap) term. We solve the scattering problem by spinor matching and derive the exact analytic expression for the transmission. The numerical results show that the drive tunes both the spin-valley-dependent propagation threshold inside the barrier and the Fabry-Pérot phase, creating controllable pass/stop bands. By varying both the laser intensity (amplitude) and the polarization shape, we show that the same junction can be switched between broadband valley filtering and resonance-selective operation, and the valley contrast remains visible in the Landauer conductance. Our findings establish an efficient route for realizing optically reconfigurable valleytronic and spintronic functionalities in MoS$_2$.
\end{abstract}
\pacs{72.80.Vp, 73.23.-b, 78.67.-n\\
	{\sc Keywords}: Monolayer MoS$_2$, Floquet engineering, electrostatic barrier, transmission, spin, valley, conductance.}
\maketitle
\section{Introduction}\label{S1}

Two-dimensional (2D) crystals have become a major platform for nanoscale electronics and photonics because their properties can be tuned efficiently using either external fields or by altering the device geometry \cite{Akinwande2019, Wang2012NNano,xia2014two,liu2019van}. Their atomic-scale thickness allows for enhanced electrostatic control, and van der Waals integration facilitates clean interfaces and versatile heterostructure engineering \cite{liu2019van, Akinwande2019}. As a consequence, simple junctions in 2D crystals often display clear quantum-transport fingerprints that can be directly exploited for device operation.
The original breakthrough in this field was the historic discovery of graphene in 2004 \cite{Novoselov2004}. It combines outstanding carrier mobility with a Dirac-like spectrum, making it a benchmark system for studies of transport in atomically thin devices \cite{geim2007rise,geim2009graphene, Neto2009}. However, the absence of a gap in this material constitutes the most important obstacle, making electrostatic confinement limited in a massless Dirac system. At normal incidence, pseudospin conservation results in perfect transmission through an electrostatic barrier, a phenomenon known as Klein tunneling \cite{katsnelson2006chiral, Neto2009}. This perfect transmission makes switching function of a solid-state device unfeasible without introducing additional symmetry-breaking perturbations \cite{Min2006, Masir2013, Mekkaoui2015}.

More recently, transition metal dichalcogenides (TMDs) monolayers of the form MX$_2$ (M = Mo, W; X = S, Se, Te) have captured a lot of attention because they successfully provide a complementary route beyond graphene \cite{chowdhury2020progress, fu20212d, yang20232d, yun2020layered}. Many of them are intrinsic semiconductors with sizable band gaps and strong spin-orbit coupling in an inversion-asymmetric environment. In particular, the monolayer semiconductor MoS$_2$ with its direct band gap shows a strong compatibility with device fabrication \cite{radisavljevic2011single, splendiani2010emerging, mak2010atomically, lv2015transition, pospischil2016optoelectronic, lembke2015single}. Its low-energy carriers reside near the two valleys $K$ and $K^{\prime}$, and intrinsic spin-orbit coupling produces valley contrast and spin splitting, leading to coupled spin- and valley degrees of freedom \cite{xiao2012coupled, Schaibley2016}. This spin-valley locking motivates valleytronics and spin-valley device concepts, where one aims to control not only the current magnitude but also its internal quantum degrees of freedom \cite{Schaibley2016,mak2014valley}.

Gate-defined barriers are the simplest adjustable junctions that are used to study spin-valley transport. These barriers turn a band spinor structure feature into an interface mode matching and generate Fabry–Pérot interference from internal reflections, yielding resonant and angle-dependent transmission \cite{cheianov2006selective, Neto2009,cheng2015transport}. In monolayer MoS$_2$, the finite mass term with the intrinsic spin-orbit coupling modifies the spinor structure and introduces propagation thresholds, then the transport through a barrier differs qualitatively from graphene and becomes channel dependent \cite{cheng2015transport}. To enhance this selectivity, a number of strategies have been explored, including velocity barriers \cite{hao2020influence}, and recent works used magnetic barriers to generate strong spin- and valley-dependent tunneling \cite{jellal2025magbarrier}. Optical control provides an additional fast tuning knob, aided by contrasting valley selection rules and valley-selective energy shifts \cite{sie2015stark, Schaibley2016}. However, the interplay between off-resonant Floquet engineering and a tunable electrostatic barrier in monolayer MoS$_2$ under elliptically polarized light remains largely unexplored. In particular, it is still unclear whether such a junction can be tuned between broadband and resonance-selective spin–valley filtering by adjusting the laser amplitude and polarization state.

{To address the aforementioned question}, we study spin- and valley-resolved transport through a single electrostatic barrier in monolayer MoS$_2$ under off-resonant elliptically polarized irradiation. The periodic drive is treated within the high-frequency Floquet framework \cite{shirley1965,li2018,wurl2018,giovannini2020,junk2020}. This yields an effective static Hamiltonian in which the light field renormalizes the mass (gap) term in a polarization-dependent way. The renormalized gap directly modifies both the propagation condition inside the barrier and the Fabry-Pérot phase accumulated in each spin-valley channel. We derive  exact analytical expression for the transmission probability and identify parameter windows where selected channels remain highly transmitting while the competing channels are strongly suppressed. The results show that an efficient field-tunable spin-valley filtering can be switched between broadband and resonance-selective operation by tuning both the laser amplitude (intensity) and its polarization state.

The paper is organized as follows. In Sec.~\ref{S2}, we introduce the driven massive-Dirac model for monolayer MoS$_2$ and derive the Floquet effective Hamiltonian and energy spectrum. In Sec.~\ref{S3}, we solve the barrier scattering problem and obtain the analytical expression 
{for} the transmission probability. Then, 
we compute the conductance for both valleys $(K,K^{\prime})$. In Sec.~\ref{S5}, we present our numerical results illustrating how the driving parameters tune the spin-valley transport windows. In Sec.~\ref{S6}, we make a comparison with graphene. Finally, in Sec.~\ref{S7}, we summarize our main results and conclude. {Appendix~\ref{appendix} presents the analytical derivations of the transmission probability}.

\section{Energy spectrum}	\label{S2}

\begin{figure}[H]
\centering
\includegraphics[scale=0.45]{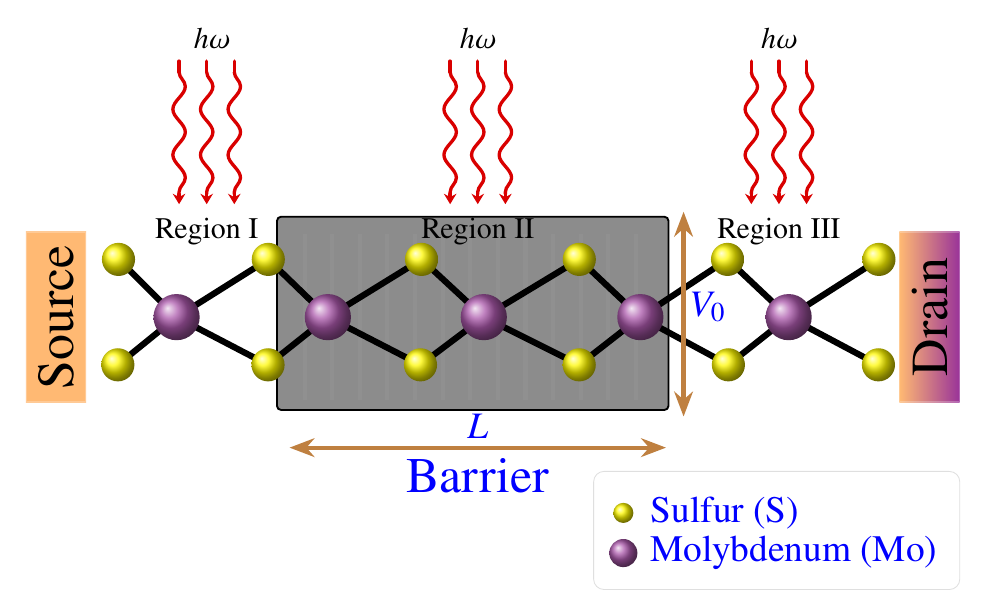}
\caption{Schematic of the uniformly irradiated monolayer MoS$_2$ junction. A rectangular electrostatic barrier of height $V_0$ and width $L$ is applied in region II, separating the source (region I) and drain (region III) leads.}
\label{fig:Tunneling}
\end{figure}

We consider a monolayer MoS$_2$ subject to an electrostatic barrier and irradiated by an off-resonant laser field, as shown schematically in Fig.~\ref{fig:Tunneling}. The scattering geometry investigated in this work consists of three distinct regions along the transport direction $x$: region I (source), region II (barrier), and region III (drain). The electrostatic potential is modeled by a rectangular barrier of height $V_0$ and width $L$. The system is further exposed to a spatially uniform off-resonant electromagnetic field with photon energy $\hbar \omega$, which does not induce real interband transitions but instead renormalizes the band structure through virtual photon processes. Within this regime, the laser field can be effectively treated using Floquet theory. This leads to modified spin- and valley-dependent parameters while preserving the stationary nature of the transport problem. Consequently, the setup reduces to a conventional three-region scattering problem, where the spin- and valley-resolved spinor wavefunctions are matched continuously at the interfaces, $x = 0$ and $x = L$.


In the low-energy approximation, the electronic properties of monolayer MoS$_2$ are governed by an effective Dirac-like Hamiltonian defined in the vicinity of the two inequivalent valleys \(K\) and \(K'\), labeled by the valley index \(\tau=\pm1\), {where $\tau=+1(-1)$ refers to the $K(K')$ valley~\cite{xiao2012coupled}}. In the presence of an electrostatic barrier and a time-periodic optical driving field, the Hamiltonian of the system can be written as
\begin{align}
H= H_0 + H_t\label{Ham1},
\end{align}
{and the two parts are given by}
\begin{align}
&H_0 =  v_F \left[ \tau \sigma_x p_x + \sigma_y p_y \right] + \frac{\Delta}{2}\sigma_z+ \frac{1-\sigma_z}{2}\lambda \tau s_z + V(x)\mathbb{I}_2 \label{Eq: H_0},\\
&
	H_t=
	-e v_F \left[\tau \sigma_x A_x(t)+\sigma_y A_y(t)\right],
	\end{align}
	where $(p_x, p_y)$ are the  momentum components. The Pauli matrices $\boldsymbol{\sigma}=(\sigma_x, \sigma_y, \sigma_z)$ describe the coupling between the conduction and valence bands. 
	The mass term $\Delta = 1.8$ {eV} breaks the inversion symmetry of the lattice, and the spin–orbit coupling introduces a spin splitting $\lambda = 0.082$ {eV} at the valence band edge, thereby modifying the electronic band structure \cite{li2013unconventional, zahid2013generic}. Here, $s_z = \pm 1$ corresponds to the two spin up and down states, $e$ represents the elementary charge, and $v_F = \frac{at}{\hbar} \approx 0.53 \times 10^6 \text{m/s}$ is the Fermi velocity, with $t$ denotes the effective hopping parameter between the Mo $d$-orbitals and S $p$-orbitals.
	The electrostatic potential defining the single rectangular barrier is modeled as
	\begin{equation}
V(x)=
\begin{cases}
	V_0, & 0 < x < L,\\
	0, & \text{otherwise},
\end{cases}
\end{equation}
where $V_0$ is the barrier height and $L$ is its width.
The light–matter interaction is introduced through the Hamiltonian $H_t$, with the vector potential of the incident light in the plane-wave representation, for an arbitrary polarization, is given by
\begin{widetext}
\begin{align}
	{\boldsymbol{A}}(t)
	= A_0 \left\{
	\left[\cos(\omega t) + \xi \cos(\omega t + \delta)\big]{\boldsymbol{e_x}}
	+ \right[-\sin(\omega t) + \xi \sin(\omega t + \delta)\big]{\boldsymbol{e_y}}
	\right\},
\end{align}
\end{widetext}
where  $A_0$ is the real amplitude of the vector potential, $\omega$ is the light frequency, and ${\boldsymbol{e_x}}$ and ${\boldsymbol{e_y}}$ are unit vectors along the $x$- and $y$-axes, respectively.
In general, $\boldsymbol{A}(t)$ describes an elliptically polarized light field. The polarization ellipse is characterized by the parameters $\xi$ and $\delta$, where $\xi$ determines the ratio of the principal axes, and $\delta$ defines the orientation of the ellipse.
Special cases include $\xi = 0$, corresponding to circular polarization, and $\xi = 1$, corresponding to linear polarization.

The Floquet formalism is used because ${\boldsymbol{A}}(t)$, associated with the laser field, causes the Hamiltonian \eqref{Ham1} to exhibit time periodicity \cite{shirley1965,li2018,wurl2018,giovannini2020,junk2020}.
{
This formalism is valid in the off-resonant regime, 
where the photon energy satisfies
$\lambda < \hbar\omega \ll \Delta$. The lower bound ensures that the photon energy exceeds the fermion energy scale set by the spin-orbit coupling, avoiding spin-flip resonances. The upper bound ensures that the photon energy remains below the band gap, thus forbidding real interband transitions. Under these conditions, the laser field acts purely through virtual-photon processes, renormalizing the band structure without inducing real transitions~\cite{Goldman2014,Mikami2016}. For the parameters used here, $\omega = 4.56\times10^{14}$~s$^{-1}$, which corresponds to a mid-infrared wavelength ($\approx4\;\mu\mathrm{m}$), giving $\hbar\omega= 0.31$~eV. This yields $\hbar\omega/\lambda \approx 3.8$, confirming that the photon energy comfortably exceeds the fermion energy scale, and $\hbar\omega/\Delta \approx 0.17$, which remains well below the band gap. Together, these ratios place the system within the off-resonant window. This is also consistent with the Floquet-engineering protocol employed experimentally in monolayer WS$_2$, where a comparable ratio $\hbar\omega/E_{\mathrm{gap}}\approx 0.155$ was reported~\cite{kobayashi2023floquet}. Within this regime, a stationary effective Floquet Hamiltonian is}
\begin{align}
H_\text{eff}=H_0+\frac{1}{\hbar\omega}
[H_{-1},H_{1}]\label{Ham2},\
\end{align}
where 
\begin{align}
	H_n=\frac{1}{T}\int_{0}^{T} H_t e^{in\omega 
		t} 
	dt,\label{eq:H_n}
\end{align}
and $T=\frac{2\pi}{\omega}$ is the period of the drive since $H(t)=H(t+T)$. Here, $H_{-1}$ and $H_{1}$ describe the virtual process of absorption and emission of a photon. 
{To evaluate the commutator $[H_{-1},H_{1}]$, we first compute $H_{\pm 1}$ 
explicitly. Substituting $\boldsymbol{A}(t)$ into~(\ref{eq:H_n}) 
and performing the time averaging over one period $T$, we find the two Hamiltonians
\begin{align}
	H_{\pm 1} = \frac{ev_FA_0}{2}
	\Big[{\tau}(1+\xi e^{\mp i\delta})\,\sigma_x 
	+ i(\mp 1\pm\xi e^{\mp i\delta})\,\sigma_y\Big].
\end{align}
Using the Pauli commutation relation 
$[\sigma_x,\sigma_y]=2i\sigma_z$, a straightforward 
calculation yields
\begin{equation}\label{eq:comm}
	[H_{-1},H_{1}] = -{\tau}(ev_FA_0)^2(\xi^2-1)\,\sigma_z,
\end{equation}
which holds for arbitrary polarization orientation $\delta$, 
as this parameter drops out exactly in the commutator.
As a result, the Hamiltonian \eqref{Ham2} can be transformed into
\begin{equation}\label{Ham3}
	H_\text{eff}=H_0 - {\tau}\frac{(ev_FA_0)^2}{\hbar\omega}(\xi^2-1) \sigma_z.
\end{equation}	


Next, we determine the eigenspinors and energies in each region. In fact, due to the translational invariance along the $y$-direction, we can decompose the spinor as
\begin{equation}
	\psi^{\tau s_z}(x, y)= \phi^{\tau s_z} (x){e}^{\mathrm{i} k_y \cdot y}.
\end{equation}
To explicitly determine $\phi^{\tau s_z}(x)$, 
{we solve} the eigenvalue equation in each region $H_\text{eff}\phi^{\tau s_z}(x)=E^{\tau s_z}(x)\phi^{\tau s_z}(x)$. We start with region I and  find the spinor solution 
\begin{widetext}
	\begin{align}
		\phi_\text{I}^{\tau s_z}(x)=\binom{1}{s\alpha^{\tau s_z} e^{i \theta^{\tau s_z}}} e^{ik_x^{\tau s_z} x}+r^{\tau s_z}\binom{1}{- s\alpha^{\tau s_z}e^{-i \theta^{\tau s_z}}}
		e^{-ik_x^{\tau s_z} x}\label{phi1} ,
	\end{align}
\end{widetext}
where $k^{\tau s_z}=\sqrt{(k_x^{\tau s_z})^2+k_y^2}$ is the wavevector modulus, $\theta^{\tau s_z}=\tan ^{-1}\frac{k_y} {k_x^{\tau s_z}}$ is the incident angle,    the parameter $\alpha^{\tau s_z}=\left(\frac{E_{1s}^{\tau s_z}-\Gamma_{ \bb \tau}}{E_{1s}^{\tau s_z}+\Gamma_{\tau}-\lambda s_z \tau}\right)^\frac{1}{2}$ is a dimensionless factor
characterizing the spinor weight in regions I and III, 
and $r^{\tau s_z}$ is the reflection coefficient. The corresponding incoming energy is 
\begin{equation}\label{ee1}
	E_{1s}^{\tau s_z}= \frac{\lambda\tau s_z}{2} +s  \hbar v_F\sqrt{(k^{\tau s_z})^2+\left(\frac{\lambda\tau s_z -2\Gamma_{\tau}}{2  \hbar v_F }\right)^2},
\end{equation}
with  
	the Floquet-renormalized mass term
	\begin{align}
		\Gamma_{\tau}= \frac{\Delta}{2}-{\tau} \frac{(ev_FA_0)^2}{\hbar\omega}(\xi^2-1)\label{Gtau},
	\end{align} 
	where
the index $s=\pm$ denotes the conduction ($s=+$) and valence ($s=-$) band solutions outside the barrier region. We can derive the wave vector component $k_x^{\tau s_z}$ as
\begin{align}
	k_x^{\tau s_z} =\tau\sqrt{\frac{1}{\hbar^2 v_F^2}\left(E_{1s}^{\tau s_z} - \Gamma_{\tau} \right)  \left(E_{1s}^{\tau s_z}- \lambda\tau s_z+ \Gamma_{\tau} \right)- k_y^2}.
\end{align}
In region III, we show
that $\phi(x)^{\tau s_z}$ has the form
\begin{align}
	\phi_\text{III}^{\tau s_z}(x)=t^{\tau s_z}\binom{1}{s\alpha^{\tau s_z}e^{i \theta^{\tau s_z}}} e^{ik_x^{\tau s_z} x},\label{phi3}
\end{align}
where $t^{\tau s_z}$ is the transmission coefficient.

Now, the solution in region II can be expressed as
\begin{widetext}
	\begin{align}
		\phi_\text{II}^{\tau s_z}(x)=a^{\tau s_z}\binom{1}{s'\beta^{\tau s_z}e^{i \varphi^{\tau s_z}}} e^{iq_x^{\tau s_z} x}+b^{\tau s_z}\binom{1}{ -s'\beta^{\tau s_z} e^{-i \varphi^{\tau s_z}}} e^{-iq_x^{\tau s_z} x},\label{phi2}
	\end{align}
\end{widetext}
where $\beta^{\tau s_z}=\left(\frac{E_{2s'}^{\tau s_z}-\Gamma_{\tau}}{E_{2s'}^{\tau s_z}+\Gamma_{\tau}-\lambda s_z \tau}\right)^\frac{1}{2}$ is a dimensionless factor denoting the spinor weight in region II, and the angle $\varphi^{\tau s_z}=\tan ^{-1}\left(\frac{k_y} { q_x^{\tau s_z}}\right)$.
The associated energy has the form
\begin{equation}\label{ee2}
	E_{2s'}^{\tau s_z}= \frac{\lambda\tau s_z}{2} + s' \hbar v_F  \sqrt{(q_x^{\tau s_z})^2+ k_y^2+\left(\frac{\lambda\tau s_z -\Gamma_{\tau}}{2v_F \hbar }\right)^2},
\end{equation}
with the index $s'=\pm$ denotes the conduction ($s'=+$) and valence ($s'=-$) band solutions inside the barrier.
This energy allows us to derive 
the  wavevector component inside the barrier 
\begin{equation}
	q_x^{\tau s_z}=\tau \sqrt{\frac{\left(E_{2s'}^{\tau s_z}-\Gamma_{ \tau}\right)\left(E_{2s'}^{\tau s_z}- \lambda\tau s_z+ \Gamma_{ \tau} \right)}{\hbar^2v_F^2} -k_y^2},
\end{equation}

\section{Tunneling transport properties
}\label{S3}

To discuss the tunneling properties, we determine the coefficients $r^{\tau s_z}$, $a^{\tau s_z}$, $b^{\tau s_z}$, and $t^{\tau s_z}$. This process can be achieved by using the continuity of the eigenspinors at the interfaces $(x = 0)$ and $(x = L)$. Imposing these boundary conditions ensures the conservation of probability current across the junction and leads to a set of coupled linear equations relating the amplitudes of the incident, reflected, and transmitted states. Solving this system yields explicit expressions for the reflection and transmission coefficients as functions of the incident energy, transverse momentum, valley index $\tau$, and spin projection $s_z$. This procedure enables a full characterization of the spin- and valley-dependent scattering processes through the barrier.
{In Appendix \ref{appendix}, we show the key intermediate steps for deriving transmission and reflection coefficients. As a result, they can be expressed as}
\begin{widetext}
	\begin{align}
		t^{\tau s_z}
		&= \frac{s s' e^{i k_x^{\tau s_z} L} \cos\theta^{\tau s_z} \cos\varphi^{\tau s_z}}
		{s s'\cos\big(q_x^{\tau s_z}L\big)\cos\theta^{\tau s_z}\cos\varphi^{\tau s_z}
			-i\sin\big(q_x^{\tau s_z}L\big)\left(\frac{(\alpha^{\tau s_z})^2+(\beta^{\tau s_z})^2}{2\alpha^{\tau s_z}\beta^{\tau s_z}}
			- s s'\sin\theta^{\tau s_z}\sin\varphi^{\tau s_z}\right)} ,\label{ttsz}
		\\
		r^{\tau s_z}
		&= \frac{\sin\big(q_x^{\tau s_z}L\big) e^{i\theta^{\tau s_z}}
			\left[s s'\sin\varphi^{\tau s_z}
			-i\frac{(\alpha^{\tau s_z})^2 e^{i\theta^{\tau s_z}}-(\beta^{\tau s_z})^2 e^{-i\theta^{\tau s_z}}}
			{2\alpha^{\tau s_z}\beta^{\tau s_z}}\right]}
		{s s'\cos\big(q_x^{\tau s_z}L\big)\cos\theta^{\tau s_z}\cos\varphi^{\tau s_z}
			-i\sin\big(q_x^{\tau s_z}L\big)\left(\frac{(\alpha^{\tau s_z})^2+(\beta^{\tau s_z})^2}{2\alpha^{\tau s_z}\beta^{\tau s_z}}
			- s s'\sin\theta^{\tau s_z}\sin\varphi^{\tau s_z}\right)} .
	\end{align}
\end{widetext}
Transmission probabilities 
be expressed as the ratio between the transmitted and incident  current densities, 
leading to the following expressions
\begin{widetext}
	\begin{align}
		T^{\tau s_z} = \frac{ \cos^2 \theta^{\tau s_z}\cos^2 \varphi^{\tau s_z}}{ \cos^2\left(q_x^{\tau s_z}L \right) \cos^2  \theta^{\tau s_z} \cos^2 \varphi^{\tau s_z}+\sin^2\left(q_x^{\tau s_z}L \right)\left(\frac{(\alpha^{\tau s_z})^2+(\beta^{\tau s_z})^2}{2\alpha^{\tau s_z}\beta^{\tau s_z}}-ss^{\prime} \sin\theta^{\tau s_z} \sin\varphi^{\tau s_z}\right)^2}.\label{eq: Transmission}
	\end{align}
\end{widetext}

The conductance provides a direct measure of the device response. Therefore, to go beyond the angle-resolved transmission, we compute the Landauer conductance for each valley $\tau$ index and spin index $s_z$ by angular averaging of the transmission probabilities $T^{\tau s_z}(\theta^{\tau s_z})$ \cite{buttiker1986four}. The associated conductances are
\begin{align}\label{GGGG}
	G^{\tau s_z} = G_0 \int_{-\pi/2}^{\pi/2} T^{\tau s_z}(\theta^{\tau s_z}) \cos\theta^{\tau s_z} \, d\theta^{\tau s_z},
\end{align}
{where $G_0 =e^2L_y/(\pi h)$ is the 
	conductance prefactor~\cite{zhai2012}, 
	with $L_y$ denoting the sample width in the $y$-direction. Since all results are presented as $G^{\tau s_z}/G_0$, the conclusions are independent of the specific value of $L_y$.} The evaluation of  \eqref{GGGG} can help us understand how spin, valley, and Floquet dressing reshape spin- and valley electron transport. As a result, we can gain useful insight into how spin- and valley-polarized currents behave in monolayer MoS$_2$. For numerical implementation, we introduce the averaged conductance as
\begin{align}
	&G_{\uparrow(\downarrow)} = \frac{G_{K\uparrow(\downarrow)} + G_{K^{\prime}\uparrow(\downarrow)}}{2},\\
	&G_{K(K^{\prime})} = \frac{G_{K(K^{\prime})\uparrow} + G_{K(K^{\prime})\downarrow}}{2}.
\end{align}

In the following, we perform a detailed numerical analysis to systematically investigate the transmission probability and conductance, with particular emphasis on their spin- and valley-resolved contributions. By varying key physical parameters such as the incident energy, angle of incidence, barrier width, and external fields, we examine how spin- and valley degrees of freedom influence the tunneling transport properties. This numerical study allows us to identify regimes of enhanced or suppressed transmission, as well as conditions leading to spin- and valley-polarized conductance, providing deeper insight into the underlying transport mechanisms.

\section{Results and discussion}\label{S5}

{The barrier height $V_0$ in our tunneling configuration can take on both positive and negative values, which create an electrostatic barrier and a potential well respectively. Such a potential well can be created experimentally by applying a local gate voltage that lowers the electrostatic potential in region II. In Fig.~\ref{fig: T_theta}, we set $V_0 = -2\Delta$, as in~\cite{hao2020influence}, to explore the potential-well regime. In this case, the angular transmission window widens, and the separation between the spin-valley channels near normal incidence becomes more visible. In the remaining figures, we set $V_0 = 0.3\Delta$, placing the system in the barrier regime where Fabry–Pérot interference dominates the transport.
	The barrier width is selected in the range $L=3$ to $10$ nm, and the electron incident energy is set to $E=1.2\Delta=2.16$ eV, in agreement with~\cite{hao2020influence} for the MoS$_2$ monolayer barrier. With this parameter setup, it is guaranteed that $E-V_0>\Gamma_{ \tau}$ throughout the entire parameter variation. As a result, the transport is governed by tunneling resonances rather than being suppressed by evanescent modes.
	Optical driving is defined through its vector-potential amplitude $A_0$, which is related to the irradiation intensity via   $I_L=\varepsilon_0\omega^2A_0^2$. The selected range of $A_0$ from $0$ to $1.2\,\mathrm{V\,fs/nm}$ represents the values reported in  experiments studying Floquet engineering in monolayer transition-metal dichalcogenides at a frequency of $4.56\times10^{14}\,\mathrm{s}^{-1}$, corresponding to a mid-infrared wavelength of approximately $4\,\mu\mathrm{m}$ \cite{kobayashi2023floquet}.
	In addition, the polarization parameter $\xi$ provides further tunability by controlling the  efficiency of Floquet dressing  relative to the field strength.  The parameters used in our simulations are therefore physically reasonable and compatible with current experiments.
}

Figure~\ref{fig: T_theta} shows the spin-valley-dependent transmission probability as a function of the incident angle. Since the potential barrier is located along the $x$-direction and the laser field alters only the mass term along $\sigma_z$, 
it does not introduce any asymmetry in $k_y$. The angular symmetry $T^{\tau s_z}(\theta^{\tau s_z}) = T^{\tau s_z}(-\theta^{\tau s_z})$ is preserved, which can be observed in all sub-figures. Furthermore, an important point is that all transmission channels have been separated into two almost identical pairs, 
{whose grouping depends on whether the laser is on. Without irradiation (panel (a)), the four channels group as $(K\uparrow, K^{\prime}\downarrow)$ and $(K\downarrow, K^{\prime}\uparrow)$, set by the spin-orbit product $\tau s_z$. Once the off-resonant drive is switched on (panels (b)-(d)), the valley-dependent gap renormalization regroups them by valley channels as $(K\uparrow, K\downarrow)$ and $(K^{\prime}\uparrow, K^{\prime}\downarrow)$,} exhibiting similar behavior within each pair but different 
behavior between pairs. 
{This pairing {in panel (a)} follows from the form of the spin-orbit term. In the Hamiltonian, it appears as $\lambda \tau s_z$. The relevant quantity is therefore the product $\tau s_z$, not the valley index and spin separately. Channels with the same value of $\tau s_z$ belong to the same spin-valley sector. In particular, $K\uparrow$ and $K'\downarrow$ both have $\tau s_z=+1$ and thus form one sector. Likewise, $K\downarrow$ and $K'\uparrow$ both have $\tau s_z=-1$ and form the other. {In contrast,} the Floquet correction in~(\ref{Ham3}) 
	{is proportional to $\tau\sigma_z$, so it adds a valley-dependent piece $\Gamma_\tau$ \eqref{Gtau}
		to the mass term. For $K$ valley, $\Gamma_+$ is increased and for $K^{\prime}$, $\Gamma_-$ is decreased. Therefore, the light dressing modifies both valleys $K$ and $K^{\prime}$ in opposite ways, and the transport is governed by the valley $\tau$ alone, rather than $\tau s_z$.}}
The oscillating lobes are Fabry-P\'erot fringes, resulting from the multiple reflections of the waves between both barrier interfaces. The transmission achieves its maximum values whenever the longitudinal phase inside the barrier satisfies the resonance condition $q_x^{\tau s_z}L\simeq n\pi$, where $n$ is an integer. 
When the irradiation is absent (panel (a)), the transmission is almost unity across a short band near normal incidence for both valleys, with only slight separation resulting in limited filtering contrast. The laser field impacts the transport by off-resonant Floquet dressing, which generates a virtual photon exchange process that adds an extra term to the mass term $\Delta$ via $\Gamma_{\tau}$ and causes a correction to the effective Hamiltonian's $\sigma_z$ term. This thereby changes $q_x^{\tau s_z}$, resulting in a new angle where the resonance condition will be satisfied. When a circularly polarized drive is switched on with $\xi=0$ and a moderate intensity $I_L=10^6\ \mathrm{W/m^2}$ ($A_0=0.71~\mathrm{V fs/nm}$) (panel (b)), the Floquet mass correction occurs, and the resonance pattern shifts. 
{Because $\Gamma_+$ is increased and $\Gamma_-$ is decreased, the two valleys respond differently. The  $K^{\prime}$ pair $(K'\uparrow, K'\downarrow)$, with the smaller effective gap $\Gamma_-$, can reach unity transmission
	near normal incidence, where both $K^{\prime}\uparrow$ and $K^{\prime}\downarrow$ form a sharp peak with $T^{K^{\prime}\uparrow\downarrow}\approx 1$, around $\theta^{\tau s_z}=0$. The $K$ pair $(K\uparrow, K\downarrow)$, with the larger gap $\Gamma_+$, is instead restricted to a smoother and lower lobe spread over a narrower angular range, with $T^{K \uparrow}\approx 0.7$-$0.8$.} To examine the role of the polarization shape, we keep the same intensity and adjust the polarization to be elliptical with $\xi=0.5$ (panel (c)). The photon exchange process becomes weaker, so the phase shift in $q_x^{\tau s_z}L$ is reduced, and the curves drift back toward the non-irradiated pattern, where the contrast between the 
{$K$ and $K^{\prime}$ valleys is diminished compared to panel (b)}. Now, we preserve this elliptical polarization (panel (d)), but we double the laser power to $I_L=2\times 10^6\ \mathrm{W/m^2}$ ($A_0=1~\mathrm{Vfs/nm}$). This makes the Floquet renormalization strong again. Thus, we recover a 
{strong valley contrast. The $K^{\prime}$ pair of channels, with the decreased gap $\Gamma_{-}$, develop sharp Fabry-P\'erot lobes that reach $T^{K^{\prime}}\approx 1$ near normal incidence and at $\theta^{\tau s_z}\approx 60^{\circ}$, separated by deep minima. The $K$ pair of channels, with the increased gap $\Gamma_+$, instead show a weaker transmission profile, with $T^{K\downarrow}\lesssim 0.7$ around $\theta^{\tau s_z}=0$.}  Fig.~\ref{fig: T_theta} therefore demonstrates that by modulating $(\xi, I_L)$, one can tune the propagation conditions and the Fabry-Pérot phase in a spin- and valley-dependent way. This gives rise to  controlled angular windows where a selected spin- and valley channel dominates the transmitted current.

\begin{figure}[ht!]
	\centering 
	\includegraphics[scale=0.55]{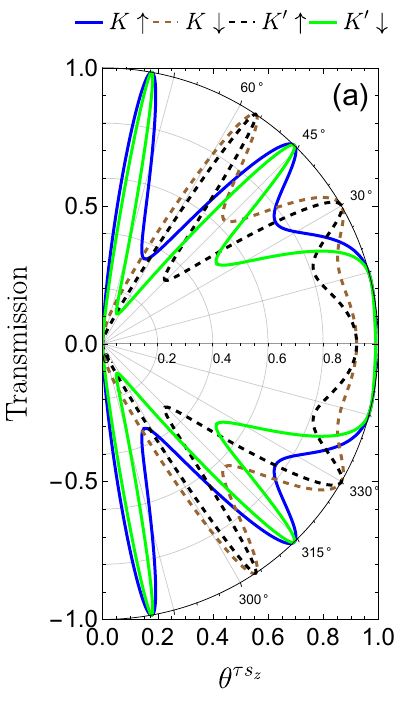}	\includegraphics[scale=0.55]{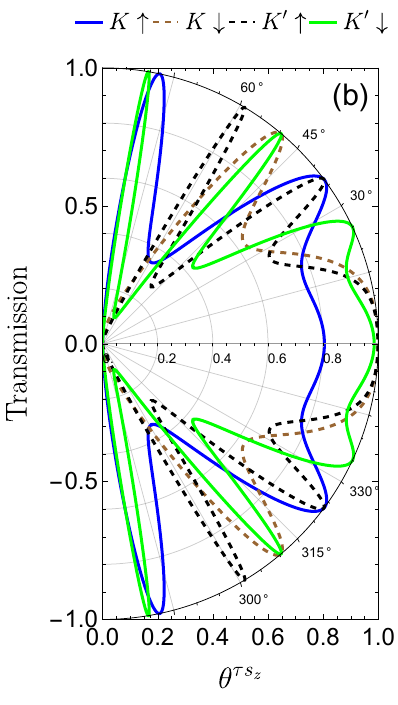}\\
	\includegraphics[scale=0.55]{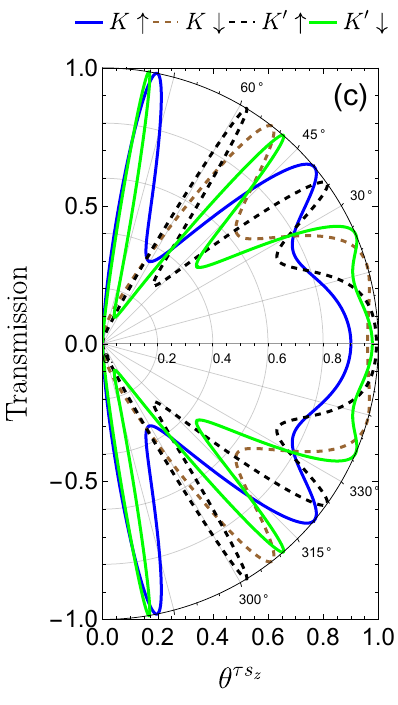}
	\includegraphics[scale=0.55]{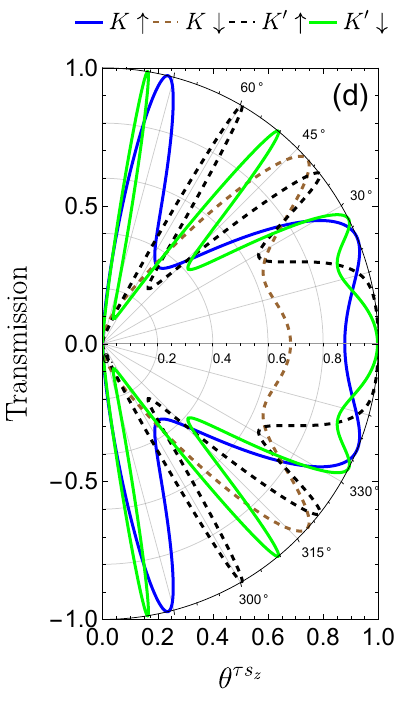}
	\caption{Spin- and valley-resolved transmission probability $T^{\tau s_z}(\theta^{\tau s_z})$ as a function of the incident angle $\theta^{\tau s_z}$ in the $K$ and $K^{\prime}$ valleys. For $E=1.2\Delta$, $V_0=-2\Delta$, $L=7~nm$. (a) No laser. (b) $\xi=0$, $I_L=10^6$ W/m$^2$. (c) $\xi=0.5$, $I_L=10^6$ W/m$^2$. (d) $\xi=0.5$, $I_L=2\times 10^6$ W/m$^2$. 
		{The radial coordinate represents the transmission $T^{\tau s_z}\in[0,1]$ and the angular coordinate represents the incident angle $\theta^{\tau s_z}\in[-90^\circ,90^\circ]$. Blue (green) solid curves denote $K\uparrow$ 
			($K^{\prime}\downarrow$), while brown (black) dashed curves denote $K\downarrow$ ($K^{\prime}\uparrow$).}}
	\label{fig: T_theta}
\end{figure}

Following the angle-resolved outcome in Fig.~\ref{fig: T_theta}, in Fig.~\ref{fig:T_L} we fix $\theta^{\tau s_z}=25^\circ$ to illustrate the transmission as a function of the barrier 
width for the purpose of scanning the Fabry-P\'erot oscillations. As indicated earlier, the resonances arise whenever the condition $q_x^{\tau s_z}L\simeq n\pi$  is met. At an intensity of $I_L=5\times10^5~\mathrm{W/m^2}$ ($A_0=0.5~\mathrm{V fs/nm}$), the fringes are closely packed and have almost the same period. For circular polarization (panel (a), $\xi=0$), the period is around $1.2$-$1.3\ \mathrm{nm}$. In both $K$ 
{channels}, the transmission remains within a high band, showing small fluctuations ($T^{K} \approx 0.8$--$1$), while the $K^{\prime}$ channels display wide stop bands with a low minimum at $T^{K^{\prime}} \approx 0.35$.  Switching to elliptical polarization (panel (b), $\xi=0.5$), both the period and the resonance width change 
slightly, suggesting a small displacement, and the period remains nearly constant. This indicates that $q_x^{\tau s_z}$ is barely changed in the low-intensity regime. As a result, the key filtering characteristic continues to be the large contrast in the depth of the oscillation between the weak modulation of $K$ channels and the deep minima of $K^{\prime}$. By increasing the intensity to $I_L=10^6~\mathrm{W/m^2}~(A_0=0.71~\mathrm{V fs/nm})$, both the period and the resonance width exhibit a significant reshaping. For circular polarization (panel (c), $\xi=0$), the oscillations stretch to a substantially larger period of around $2.7~\text{to}~2.8~\mathrm{nm}$, 
{because $\Gamma_+$ is increased and the  wavevector $q_x^{K}$ inside the barrier is consequently reduced. The $K^{\prime}$ oscillations instead keep a smaller period, because $\Gamma_-$ is decreased and $q_x^{K^{\prime}}$ is consequently increased. The peaks of both valleys still reach unity. Between resonances, however, both valleys develop deep minima of comparable depth ($T^{K}\simeq T^{K^{\prime}}\simeq 0.3$--$0.35$), so the strong contrast between $K$ and $K^{\prime}$ seen in panels (a) and (b) is largely absent in this regime.} For elliptical polarization at the same high intensity (panel (d), $\xi=0.5$), the period decreases again to $1.6$-$1.7~\mathrm{nm}$, and the resonances become 
{stronger}, producing a 
{wide range of width values for which} the $K$ channels keep relatively high minima ($T^{K}\simeq0.65$), whereas the $K^{\prime}$ channels still dip much lower (typically $T^{K^{\prime}}\simeq$
{$0.35$}).  
Fig.~\ref{fig:T_L} shows that $I_L$ primarily controls the stretching of the period and the sharpness of the resonances, while increasing $\xi$ tends to compress the period and expand the pass bands. This tunability allows one to switch between resonance-selective and broadband spin- and valley-filtering by choosing $L$ in a stop band of the suppressed valley.

\begin{figure}[ht!]
	\centering 
	\includegraphics[scale=0.465]{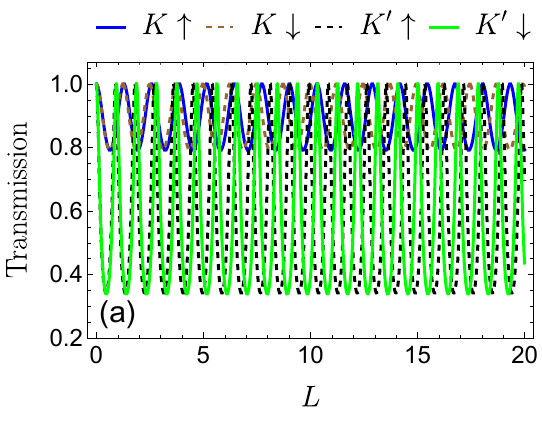}	\includegraphics[scale=0.465]{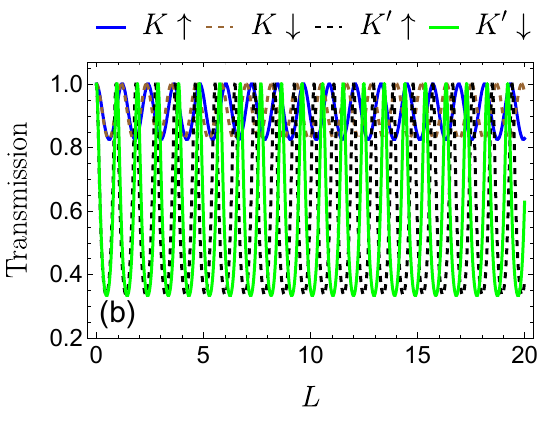}\\
	\includegraphics[scale=0.465]{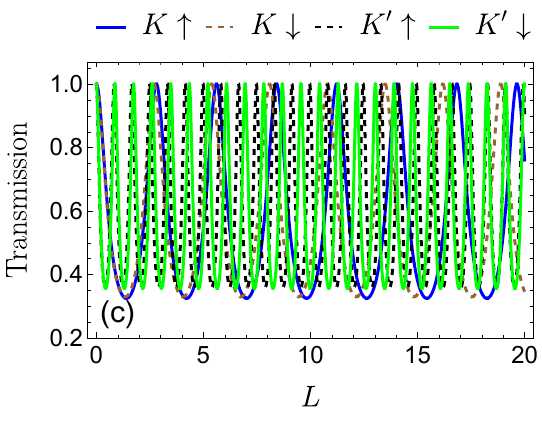}
	\includegraphics[scale=0.465]{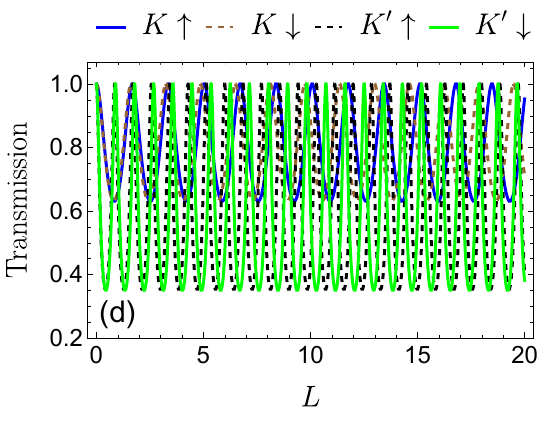}
	\caption{Spin- and valley-resolved transmission probability $T^{\tau s_z}(L)$ as a function of the barrier width $L$ in the $K$ and $K^{\prime}$ valleys. For $E=1.2\Delta$, $V_0=0.3\Delta$, $\theta^{\tau s_z}=25^\circ$. (a)  $\xi=0$, $I_L=5\times 10^5$ W/m$^2$. (b) $\xi=0.5$, $I_L=5\times 10^5${~W/m$^2$}. (c) $\xi=0$, $I_L=10^6$ W/m$^2$. (d) $\xi=0.5$, $I_L=10^6$ W/m$^2$. Blue (green) solid curves denote $K\uparrow$ ($K^{\prime}\downarrow$), while brown (black) dashed curves denote $K\downarrow$ ($K^{\prime}\uparrow$).}
	\label{fig:T_L}
\end{figure}

\begin{figure}[ht!]
	\centering 
	\includegraphics[scale=0.465]{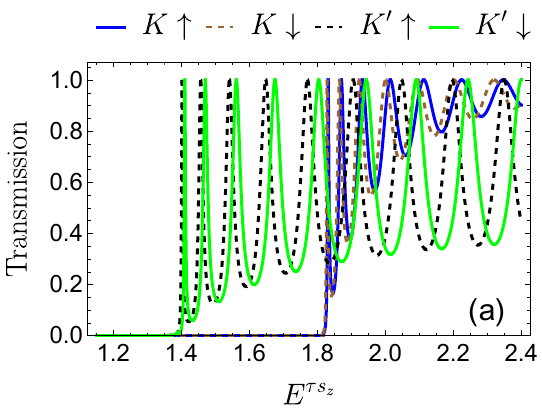}
	\includegraphics[scale=0.465]{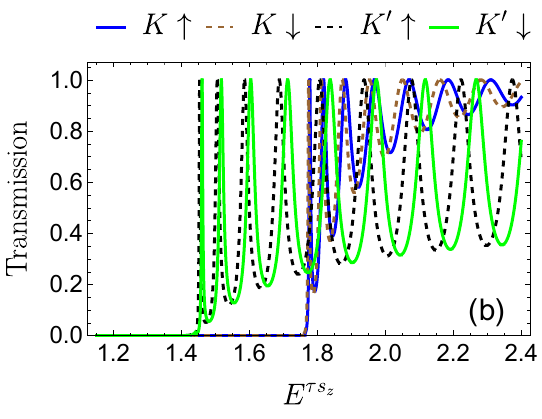}\\
	\includegraphics[scale=0.465]{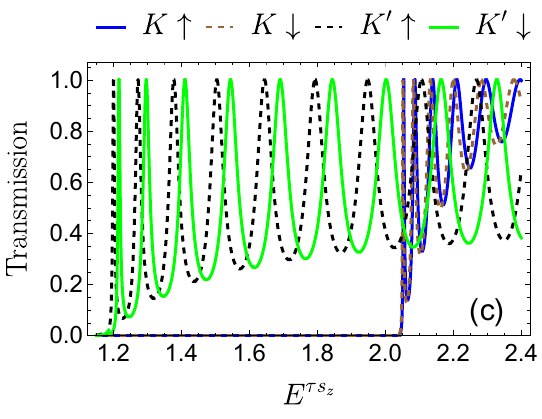}
	\includegraphics[scale=0.465]{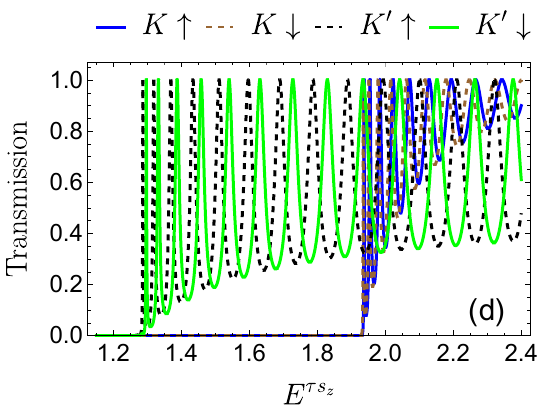}
	\caption{Spin- and valley-resolved transmission probability $T^{\tau s_z}(E^{\tau s_z})$ as a function of the incident energy $E^{\tau s_z}$ in the $K$ and $K^{\prime}$ valleys. For $V_0=0.3\Delta$, $L=7~nm$, $\theta^{\tau s_z}=25^\circ$. (a) $\xi=0$, $I_L=5\times 10^5$ W/m$^2$. (b) $\xi=0.5$, $I_L=5\times 10^5$ W/m$^2$. (c) $\xi=0$, $I_L=10^6$ W/m$^2$. (d) $\xi=0.5$, $I_L=10^6$ W/m$^2$.  Blue (green) solid curves denote $K\uparrow$ ($K^{\prime}\downarrow$), while brown (black) dashed curves denote $K\downarrow$ ($K^{\prime}\uparrow$).}
	\label{fig:T_E}
\end{figure}

We have shown in Figs.~\ref{fig: T_theta} and \ref{fig:T_L} that the filter is set by the phase accumulated inside the barrier $q_x^{\tau s_z}L$. We now scan the same physics in energy at fixed $L=7$~nm and $\theta^{\tau s_z}=25^\circ$. Fig.~\ref{fig:T_E} indicates  that the transmission is activated only when the mode inside the barrier starts propagating, i.e., $q_x^{\tau s_z}$ is real. This defines a threshold energy $E_{\rm th}^{\tau s_z}$, below which 
$q_x^{\tau s_z}$ is imaginary, the wave is evanescent, and the transmission is strongly blocked. This {onset} energy 
varies from panel to panel and presents a clear signature of the photon-exchange process in the Floquet regime, which renormalizes the mass term through $\Gamma_\tau$} and consequently modifies the effective gap. At an intensity of $I_L=5\times10^5$~W/m$^2~(A_0=0.5~\mathrm{V fs/nm})$, the dressing is moderate, and the onset 
{already splits by valley in panels (a) and (b): the $K$ channels open at $E_{\rm th}^{K}\approx 1.81$~eV in panel (a) and $E_{\rm th}^{K}\approx 1.76$~eV in panel (b), while the $K^{\prime}$ channels are already active from $E_{\rm th}^{K^{\prime}}\approx 1.40$~eV in panel (a) and $E_{\rm th}^{K^{\prime}}\approx 1.45$~eV in panel (b)}. At the higher intensity, $I_L=10^6$~W/m$^2~(A_0=0.71~\mathrm {V fs/nm})$, the threshold becomes strongly 
{valley-split. For circular polarization (panel (c), $\xi=0$), the increased gap $\Gamma_+$ pushes the $K$ threshold up to $E_{\rm th}^{K}\approx 2.05$~eV, while the decreased gap $\Gamma_-$ pulls the $K^{\prime}$ threshold down to $E_{\rm th}^{K^{\prime}}\approx 1.20$~eV. For elliptical polarization (panel (d), $\xi=0.5$), the splitting is smaller: $E_{\rm th}^{K}\approx 1.93$~eV and $E_{\rm th}^{K^{\prime}}\approx 1.30$~eV. This valley asymmetry is expected because $\Gamma_\tau$ increases the effective gap for $\tau=+1$ ($K$) and decreases it for $\tau=-1$ ($K^{\prime}$), shifting their respective onset energies in opposite directions.}
For an energies higher than $E_{\rm th}^{\tau s_z}$,  changes in $E^{\tau s_z}$ alter the phase inside the barrier, $q_x^{\tau s_z}(E^{\tau s_z})L$, causing the transmission to oscillate and reach maxima when the resonance condition is satisfied. Just above the onset, the peaks are narrower, and they broaden as $E^{\tau s_z}$ increases because $q_x^{\tau s_z}$ rises with $E^{\tau s_z}$. {In panels (a) and (b),} for the $K$ valley, both channels generate high-transmission oscillations with slight modulation ($T^{K}\gtrsim 0.8$ and approach unity for $E\gtrsim 2.2$~eV). However, the $K^{\prime}$ channels exhibit deep stop bands, with minima $T^{K^{\prime}}\sim 0.25$-$0.3$ 
and down to $\sim 0.1$ near the onset. 
{In panel (c), the large valley splitting of the threshold creates an exclusive window $1.20\lesssim E^{\tau s_z}\lesssim 2.04$~eV in which only $K^{\prime}$ transmits; above $2.04$~eV both valleys are active and $K$ oscillates between $T^{K}\approx 0.4$ and unity. In panel (d), the window narrows to $1.30\lesssim E^{\tau s_z}\lesssim 1.95$~eV, with $K^{\prime}$ oscillating throughout and $K$ building up to $T^{K}\gtrsim 0.4$ above its onset.} We stress that $(\xi, I_L)$ controls both the opening 
{energies $E_{\rm th}^{K}$ and $E_{\rm th}^{K^{\prime}}$ through the valley-asymmetric Floquet gap renormalization $\Gamma_\tau$} and the pass/stop-band positions through $q_x^{\tau s_z}(E^{\tau s_z})L$, providing an efficient mechanism for spin- and valley filtering.

\begin{figure}[ht!]
	\centering 
	\includegraphics[scale=0.465]{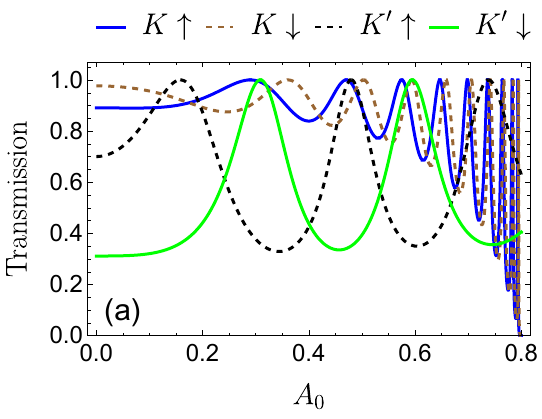}
	\includegraphics[scale=0.465]{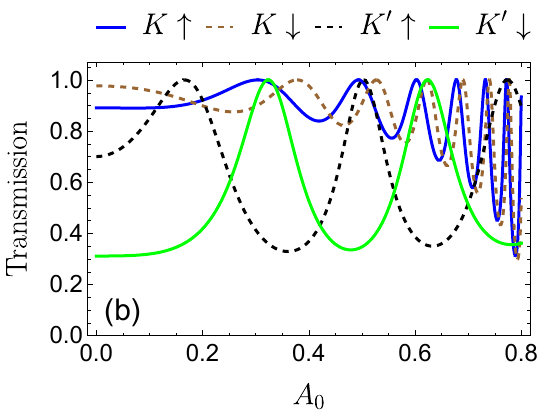}\\
	\includegraphics[scale=0.465]{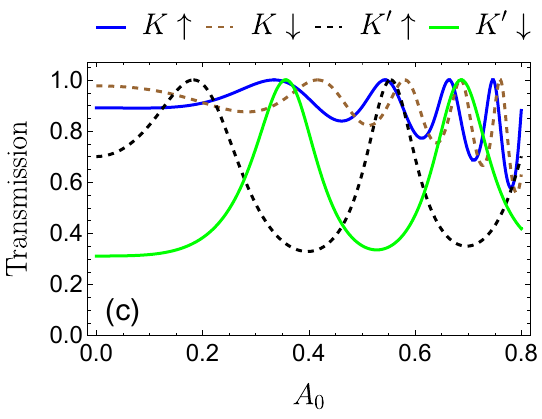}
	\includegraphics[scale=0.465]{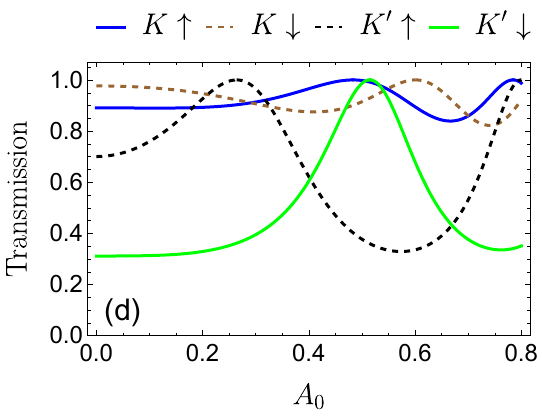}
	\caption{Spin- and valley-resolved transmission probability $T^{\tau s_z}(A_0)$ as a function of the laser-field amplitude $A_0$ in the $K$ and $K^{\prime}$ valleys. For $E=1.2\Delta$, $V_0=0.3\Delta$, $L=10~\mathrm{nm}$, and $\theta^{\tau s_z}=25^\circ$. (a) $\xi=0$. (b) $\xi=0.3$. (c) $\xi=0.5$. (d) $\xi=0.8$. Blue (green) solid curves denote $K\uparrow$ ($K^{\prime}\downarrow$), while brown (black) dashed curves denote $K\downarrow$ ($K^{\prime}\uparrow$).}
	\label{fig:T_A0}
\end{figure}

\begin{figure}[ht!]
	\centering 
	\includegraphics[scale=0.465]{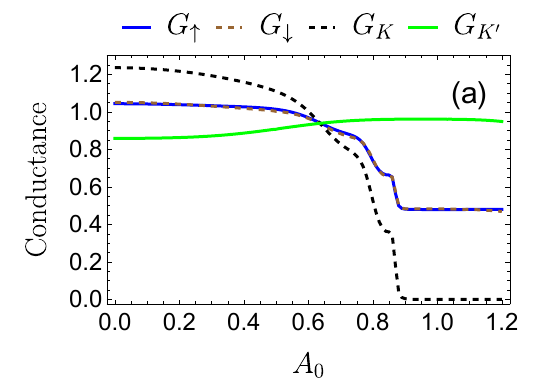}
	\includegraphics[scale=0.465]{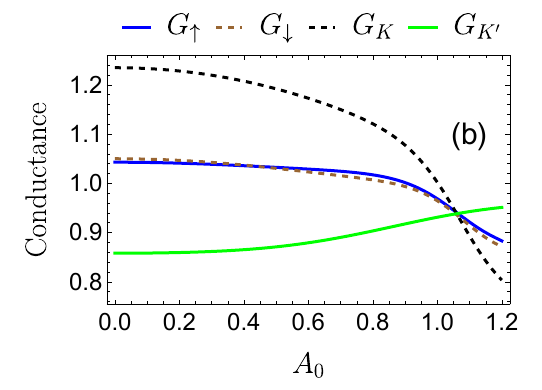}
	\caption{Spin- and valley-averaged conductance $G^{\tau s_z}(A_0)/G_0$ as a function of the laser amplitude $A_0$ in the $K$ and $K^{\prime}$ valleys For $E=1.2\Delta$, $V_0=0.3\Delta$, $L=3~\mathrm{nm}$. (a) $\xi=0$ (circular polarization). (b) $\xi=0.8$ (close to linear).  Blue (green) solid curves show $G_{\uparrow}$ ($G_{K^{\prime}}$), while brown  (black) dashed curves show $G_{\downarrow}$ ($G_{K}$).}
	\label{fig:G_A0}
\end{figure}

To complete the control picture, we show in Fig.~\ref{fig:T_A0} the transmission as a function of the irradiation amplitude $A_0$, {which acts} as an optical gate for the spin- and valley filtering. In the off-resonant regime, virtual photon exchange renormalizes the mass term through $\Gamma_{\tau}$. Therefore, increasing $A_0$ shifts both $q_x^{\tau s_z}$ and the Fabry-P\'erot phase $q_x^{\tau s_z}L$. In the weak-{drive} regime ($A_0 \lesssim 0.2$--$0.25$), both valleys exhibit strongly different responses. The transmission for both $K$ channels remains close to a transparent plateau ($T^{K} \simeq 0.9$--$1$), while it is low for $K^{\prime}$ channels ($T^{K^{\prime}\downarrow} \simeq 0.3$,
{while $T^{K^{\prime}\uparrow}$ starts from $\simeq 0.7$ at $A_0=0$}). As $A_0$ rises, the $K^{\prime}$ channels undergo resonant openings where the phase condition is met, and the transmission approaches unity. For $\xi = 0$ and $\xi = 0.3$ (panels (a) and (b)), these apertures develop 
{a first $K^{\prime}\uparrow$ peak near $A_0\simeq 0.15$ and a first $K^{\prime}\downarrow$ peak near $A_0\simeq 0.30$}, and then evolve into a 
fringe pattern for $A_0 \gtrsim {0.65}$ with a decreasing spacing. In the $K^{\prime}$ valley, the transmission can drop to low values down to 
{($\simeq 0.3$ for $K^{\prime}\uparrow$)}, whereas the $K$ channels remain highly transparent {on a plateau ($T^{K}\gtrsim 0.7$) up to $A_0\simeq 0.55$, beyond which they also develop Fabry-P\'erot oscillations and, in panel (a), $T^{K\uparrow}$ collapses sharply toward zero near $A_0\simeq 0.75$.} The results reveal that setting $A_0$ provides an optical 
{phase-tunable} spin-valley filtering. On the other hand, increasing the polarization $\xi$ leads to a weak photon-exchange mechanism. For $\xi = 0.5$ (panel (c)), the fringes are less compressed, and only a few resonances survive in the range $0.5 \lesssim A_0 \lesssim 0.8$,  the filtering contrast therefore is reduced. At $\xi = 0.8$ (panel (d)), the response is dominated by a small number of broad resonant features,
thus indicating a wide pass/stop band instead of rapid oscillations.  Fig.~\ref{fig:T_A0} indicates that $A_0$ regulates the strength of Floquet dressing, which substantially modulates the Fabry-P\'erot accumulation phase inside the barrier, while $\xi$ modulates the density of the fringes. This makes the same barrier act either as a narrow-band, high-contrast optical filter at small $\xi$ or as a more robust broadband selector at larger $\xi$, 
{where the valley-asymmetric Floquet dressing makes the $K$ channels develop slower oscillations set by the increased gap $\Gamma_+$ while the $K^{\prime}$ channels develop faster oscillations set by the decreased gap $\Gamma_-$}. {Therefore, this finding allows us to draw an explicit connection between the filtering regime and the Floquet gap renormalization. Increasing $A_0$ raises the renormalized gap $\Gamma_{+}$, {of the $K$ valley and lowers $\Gamma_{-}$ of the $K^{\prime}$ valley}, which directly shifts the longitudinal wavevector $q_x^{\tau s_z}$ inside the barrier and consequently modifies the Fabry-P\'erot accumulation phase $q_x^{\tau s_z}L$, {in opposite directions for the two valleys}. When this phase shift is small, the $K^{\prime}$ stop bands are wide, and the filter operates in the broadband regime. When the phase shift becomes large and varies rapidly with $A_0$, the stop bands narrow and the filter enters the resonance-selective regime. The critical amplitude separating these two regimes is $A_0^{\rm c}\simeq 0.55$~V fs/nm for $\xi=0$ and $L=10$~nm, at which the Floquet-induced correction to $\Gamma_{\tau}$ reaches approximately $0.28$~eV 
	{in absolute value with respect to} the gap value $\Delta/2=0.9$~eV. Beyond this amplitude, the phase sensitivity increases sharply, producing the dense fringe pattern seen in Fig.~\ref{fig:T_A0}. 
	{At $A_0\simeq 0.75$~V fs/nm in panel (a), $\Gamma_+$ for the $K$ valley grows large enough to drive $q_x^{K}$ toward the evanescent regime, producing the sharp collapse of $T^{K\uparrow}$ visible at the right edge of the panel and consistent with the valley-selective switch-off of $G_K$ shown in Fig.~\ref{fig:G_A0}.}}

\begin{figure}[ht!]
	\centering 
	\includegraphics[scale=0.465]{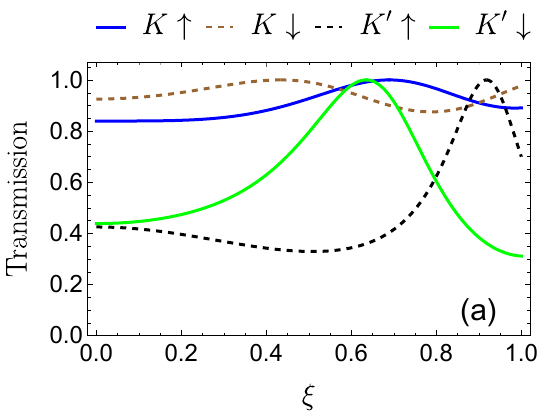}
	\includegraphics[scale=0.465]{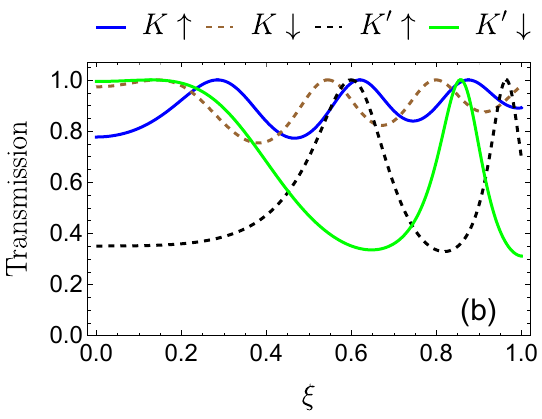}\\
	\includegraphics[scale=0.465]{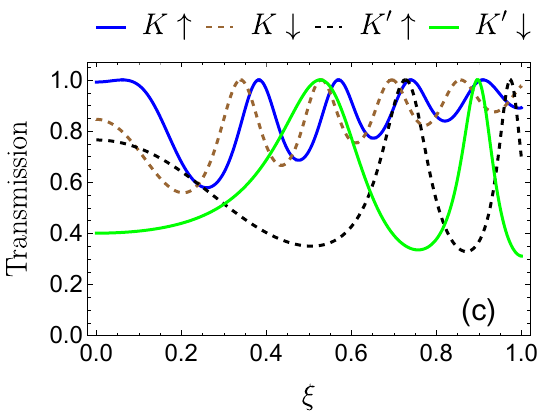}
	\includegraphics[scale=0.465]{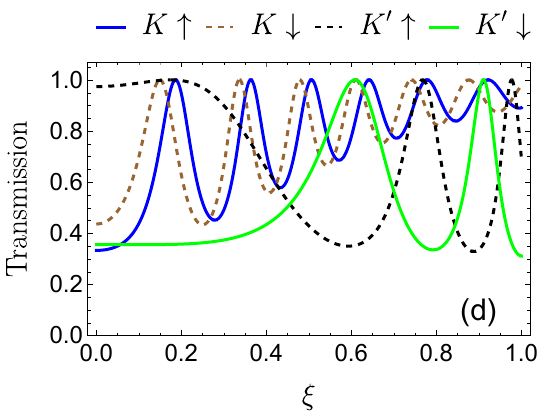}
	\caption{Spin- and valley-resolved transmission probability $T^{\tau s_z}(\xi)$ as a function of the polarization shape $\xi$ in the $K$ and $K^{\prime}$ valleys. For $E=1.2\Delta$, $V_0=0.3\Delta$, $L=10~\mathrm{nm}$, and $\theta^{\tau s_z}=25^\circ$. (a) $A_0=0.4~\mathrm{V fs/nm}$. (b) $A_0=0.6~\mathrm{V fs/nm}$. (c) $A_0=0.7~\mathrm{V fs/nm}$. (d) $A_0=0.75~\mathrm{V fs/nm}$. Blue (green) solid curves denote $K\uparrow$ ($K^{\prime}\downarrow$), while brown (black) dashed curves denote $K\downarrow$ ($K^{\prime}\uparrow$).}
	\label{fig:T_xi}
\end{figure}

Figure~\ref{fig:T_xi} displays the transmission as a function of the polarization shape $\xi$. In the off-resonant regime, the virtual photon exchange intensity varies with the variation of $\xi$. Thus, the Floquet mass correction in $\Gamma_{\tau}$ reaches its highest  value for $\xi=0$ and decreases continuously toward $\xi \to 1$. As discussed previously, this correction shifts $q_x^{\tau s_z}$ and modifies the phase $q_x^{\tau s_z}L$. For the weakest drive (panel (a)), both $K$ channels stay highly transmitted ($T^{K}\simeq 0.85$--$1$), whereas the $K^{\prime}$ channels show large transmission modulations, 
{both starting from $T^{K^{\prime}}\simeq 0.4$ at $\xi=0$, with $K^{\prime}\downarrow$ rising to a unity  peak near $\xi\simeq 0.65$ and then falling to $T^{K^{\prime}\downarrow}\simeq 0.3$ at $\xi=1$, and with $K^{\prime}\uparrow$ exhibiting a shallow minimum $T^{K^{\prime}\uparrow}\simeq 0.3$ near $\xi\simeq 0.5$ and a peak $T^{K^{\prime}\uparrow}\simeq 1$ near $\xi\simeq 0.9$}. With stronger driving (panel (b)), $\xi$ sweeps the phase through several resonances where $K^{\prime}$ develops repeated pass bands (peaks near $\xi\simeq$  {0., $0.6$, and $0.85$, $0.95$} with $T^{K^{\prime}}\approx 1$) separated by deep minima 
with a transmission dropping  to $T^{K^{\prime}}\sim {0.3}$. However, the current remains dominated by $K$ channels ($T^{K}\gtrsim 0.75$). At larger intensities (panels (c) and (d)), the pattern becomes resonance-selective {for both valleys.} Here, the $K^{\prime}$ stop bands exhibit 
{deep minima down to $T^{K^{\prime}}\simeq 0.3$ in both panels}. Therefore, the pass bands narrow, so that slight changes in $\xi$ flip $K^{\prime}$ between a practically open and an entirely blocked channel. In comparison, the $K$ valley transmission is relatively constant and remains near unity for $\xi\gtrsim 0.7$, while both valleys oscillate at large amplitudes. Fig.~\ref{fig:T_xi} illustrates that $\xi$ acts as a polarization gate through the Floquet mass correction $\propto(\xi^2-1)$, which modifies $q_x^{\tau s_z}$ and consequently the phase $q_x^{\tau s_z}L$. As $\xi$ is swept, the $K^{\prime}$ channels alternate between pass and stop bands, whereas the $K$ channels stay at high-transmission plateau, {at small $A_0$ and develop their own oscillations at the largest amplitudes.} Increasing $A_0$ raises phase sensitivity to $\xi$, narrowing the $K^{\prime}$ pass/stop windows and making filtering more resonance-selective.

\begin{figure}[ht!]
	\centering 
	\includegraphics[scale=0.465]{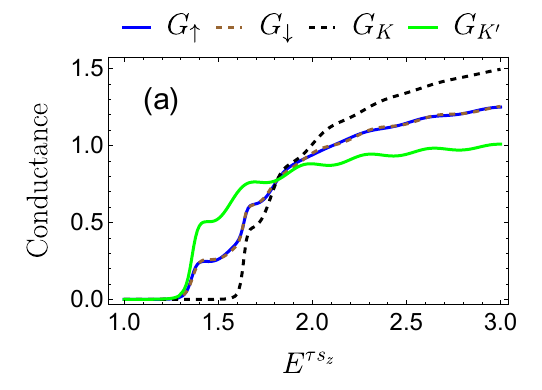}
	\includegraphics[scale=0.465]{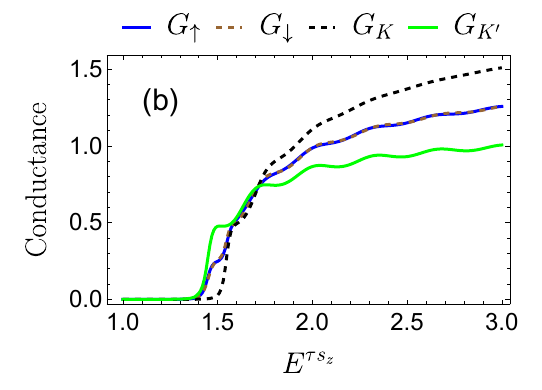}\\
	\includegraphics[scale=0.465]{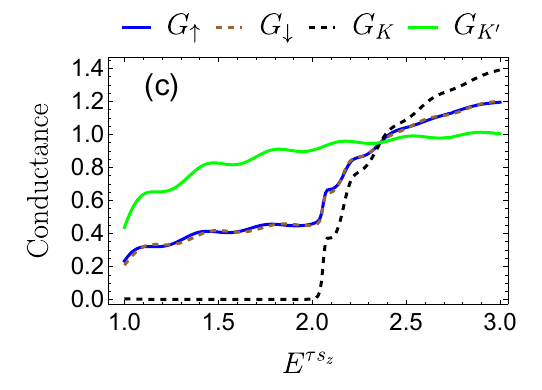}
	\includegraphics[scale=0.465]{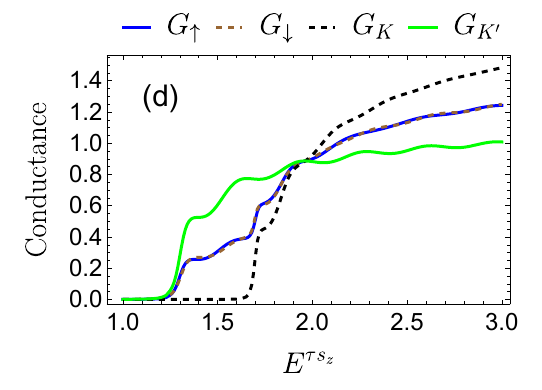}
	\caption{Spin- and valley-averaged conductance $G_{\tau s_z}(E^{\tau s_z})/G_0$ as a function of the incoming energy $E^{\tau s_z}$ for $V_0=0.3\Delta$ and $L=3~\mathrm{nm}$.
		(a) $A_0=0.4~\mathrm{V fs/nm}$ and $\xi=0$ (circular polarization).
		(b) $A_0=0.4~\mathrm{V fs/nm}$ and $\xi=0.8$ (close to linear).
		(c) $A_0=0.8~\mathrm{V fs/nm}$ and $\xi=0$ (circular).
		(d) $A_0=0.8~\mathrm{V fs/nm}$ and $\xi=0.8$ (close to linear).
		Blue (green) solid curves show $G_{\uparrow}$ ($G_{K^{\prime}}$), while brown  (black) dashed curves show $G_{\downarrow}$ ($G_{K}$).}
	\label{fig:G_E}
\end{figure}

Figure~\ref{fig:G_E} shows the spin- and valley-resolved conductance obtained from the Landauer angular average of the transmission provided in Eq.~(\ref{GGGG}). Because $G^{\tau s_z}(E^{\tau s_z})$ integrates $T^{\tau s_z}(E^{\tau s_z},\theta^{\tau s_z})$ over the whole range of incident angles $\theta^{\tau s_z}$, it suppresses most of the Fabry-P\'erot fringes shown in Fig.~\ref{fig:T_E}. However, it preserves two important fingerprints for  device operation: a distinct onset energy and a robust valley contrast. For all panels, the conductance remains zero below the threshold $E_{\rm th}^{\tau s_z}$, which is aligned with the propagation condition discussed above in Fig.~\ref{fig:T_E} for the transmission. We have seen that for an incoming energy lower than a value $E_{\rm th}^{\tau s_z}$, the longitudinal wave vector inside the barrier is imaginary for most of  incident angles. Therefore, the tunneling is dominated by evanescent modes. While once the energy exceeds $E_{\rm th}^{\tau s_z}$, a propagating state immediately pens inside the barrier region for a finite angular window, resulting in a fast growth of $G^{\tau s_z}(E^{\tau s_z})$ over this range. While the visible weak residual resonances are the trace of the Fabry–P\'erot phase $q_x^{\tau s_z}(E^{\tau s_z})L$ produced by the angle-averaging. Across all panels, the filtering is mainly valley selective. 
{It is shown that the two valley conductances $G_{K}$ (black dashed) and $G_{K^{\prime}}$ (green solid) are strongly separated by the valley-asymmetric Floquet gap renormalization $\Gamma_\tau$. At low energies, $G_{K^{\prime}}$ rises first because the decreased gap $\Gamma_-$ pulls the $K^{\prime}$ threshold below the $K$ threshold, while $G_{K}$ remains near zero until the higher onset $E_{\rm th}^{K}$ is reached. Above $E_{\rm th}^{K}$, $G_{K}$ grows rapidly and eventually exceeds $G_{K^{\prime}}$ at sufficiently high energy,  consistent with the behavior revealed earlier in Fig.~\ref{fig:T_E}.} 
Moreover, $G_{\uparrow}$ and $G_{\downarrow}$ practically overlap,  confirming that the outcome is mostly valley selective for this set of parameters.   
{The weak spin contrast, namely $G_{\uparrow}\approx G_{\downarrow}$, originates from the intrinsic spin-valley structure of the MoS$_2$ Hamiltonian. Since the spin-orbit interaction enters through the product $\lambda\tau s_z$, the transport channels are organized by the combined index $\tau s_z$. As a result, the contribution  of one spin  in one valley is largely compensated by the corresponding contribution from the opposite valley when the conductance is averaged over valleys. Therefore, even when the valley contrast is strong, the net spin-resolved conductances remain close. The Floquet dressing modifies the propagation thresholds and Fabry-P\'erot resonances through the renormalized gap term, but it does not remove this underlying $\tau s_z$ structure. Consequently, the compensation between spin-up and spin-down contributions remains intact. Enhancing spin selectivity requires reducing this compensation. A direct route is to introduce a magnetic field or a ferromagnetic proximity exchange field, which induces a spin-dependent energy shift and separates the propagation conditions for spin-up and spin-down channels within each valley, as reported in MoS$_2$ junctions~\cite{hao2020influence,jellal2025magbarrier}.} 
The optical control takes place through the shift of $E_{\rm th}^{\tau s_z}$. For $A_0=0.4\ \mathrm{V fs/nm}$, 
{the thresholds split by valley: in panel (a) ($\xi=0$, circular) we read $E_{\rm th}^{K}\simeq 1.6$~eV and $E_{\rm th}^{K^{\prime}}\simeq 1.30$~eV, while in panel (b) ($\xi=0.8$, nearly linear) the splitting decreases to $E_{\rm th}^{K}\simeq 1.50$~eV and $E_{\rm th}^{K^{\prime}}\simeq 1.40$~eV, consistent with a weaker Floquet mass correction as $\xi$ approaches 1.} When the field is increased to $A_0=0.8\ \mathrm{V fs/nm}$, the dressing is stronger in panel (c) ($\xi=0$, circular), and the onset 
{of $K$ moves up to $E_{\rm th}^{K}\simeq 2.05$~eV while the onset of $K^{\prime}$ falls below $1.0$~eV, opening a wide energy window in which only $K^{\prime}$ contributes to the conductance and greatly enhancing the valley separation.} 
However, in panel (d) for $\xi=0.8$, the dressing is reduced and pulls the 
{thresholds back to $E_{\rm th}^{K}\simeq 1.65$~eV and $E_{\rm th}^{K^{\prime}}\simeq 1.25$~eV}, resulting in an intermediate valley contrast. It is clear that $A_0$ essentially sets the activated energy scale, whereas $\xi$ provides a polarization gate for the valley-filter performance.

Fig.~\ref{fig:G_A0} illustrates the conductance $G^{\tau s_z}(A_0)/G_0$ as a function of the laser amplitude $A_0$ for two different polarization shapes. 
{Both panels show a predominantly valley-selective response with $G_{\uparrow}\approx G_{\downarrow}$, so the spin contrast remains weak. However, in contrast with Fig.~\ref{fig:G_E}, the relative ordering of the two valleys is now $A_0$-dependent: $G_{K}>G_{K^{\prime}}$ at small $A_0$, but the laser-induced gap renormalization $\Gamma_\tau$ increases the gap of $K$ and decreases that of $K^{\prime}$, so as $A_0$ increases the two curves cross and $G_{K^{\prime}}$ eventually exceeds $G_{K}$.} For  circular polarization $\xi=0$ in panel (a), the system begins with a conducting regime with a clear valley selectivity. For small values of $A_0$, $G_{K}\simeq 1.23$ and $G_{K^{\prime}}\simeq 0.86$, whereas $G_{\uparrow}\approx G_{\downarrow}\simeq 1.05$. When $A_0$ increases, 
{$G_{K}$ decreases gradually until to $A_0\simeq 0.7$ and then collapses rapidly, falling to the off-state at $A_0\simeq 0.9$, while $G_{K^{\prime}}$ stays nearly constant around $G_{K^{\prime}}\simeq 0.95$ and the spin-resolved conductances stabilize at $G_{\uparrow}\approx G_{\downarrow}\simeq 0.5$.} 
{This provides a valley-selective optical switch: the Floquet gap renormalization drives the $K$ channel into an evanescent regime for most incident angles and switches it off, while $K^{\prime}$, whose gap is decreased by the same drive, remains conducting. Thus $A_0$ acts as an on/off gate for the $K$ valley and converts the junction into an essentially $K^{\prime}$-only conductor at strong drive.} For nearly linear polarization $\xi=0.8$ in panel (b), the suppression is smoother. Here, the  Floquet correction is weaker, so the conductance decreases continuously and does not fully vanish in the plotted range. Even at $A_0=1.2$, it stays finite, with $G_{K}\approx 0.80$ and $G_{K^{\prime}}\approx {0.95}$,  moreover $G_{\uparrow}\approx G_{\downarrow}$. We conclude that increasing $\xi$ turns the laser control from a switching mechanism to a continuous tuning knob while preserving the key functionality. 
{At small $A_0$, the current is dominated by the $K$ valley, but as $A_0$ increases the valley-asymmetric Floquet dressing causes the two valley conductances to cross, and the $K^{\prime}$ valley becomes the dominant channel.}

\section{MoS$_2$ and  graphene}\label{S6}

Graphene and monolayer MoS$_2$ are both two-dimensional Dirac materials, since their low-energy carriers reside near the two inequivalent valleys $K$ and $K^{\prime}$. However, their intrinsic energy scales are very different. Graphene is effectively gapless, and its intrinsic spin-orbit coupling is extremely weak, while monolayer MoS$_2$ is a direct-gap semiconductor with strong spin-orbit coupling generated by broken inversion symmetry \cite{splendiani2010emerging,mak2010atomically,xiao2012coupled, Schaibley2016}. This difference is already encoded in our general transmission formula, Eq.~(\ref{eq: Transmission}), through the spinor-matching factors $\alpha^{\tau s_z}$ and $\beta^{\tau s_z}$. In graphene, the mass term is absent, and  the two pseudospin components of a propagating state, therefore, have equal weight, causing the matching parameters to reduce to $\alpha^{\tau s_z}=\beta^{\tau s_z}=1$ (outside and inside the barrier) \cite{Neto2009,katsnelson2006chiral}. Substituting $\alpha^{\tau s_z}=\beta^{\tau s_z}=1$ into Eq.~(\ref{eq: Transmission}) directly yields the standard graphene barrier transmission. It is controlled only by the incidence/refraction angles and the Fabry-P\'erot phase. In particular, it reproduces perfect transmission at normal incidence (Klein tunneling) \cite{katsnelson2006chiral,Neto2009}. For MoS$_2$, the finite mass term and intrinsic spin-orbit splitting make $\alpha^{\tau s_z}$ and $\beta^{\tau s_z}$ depart from unity and become channel-dependent through $\tau s_z$ \cite{xiao2012coupled, Schaibley2016}, while the off-resonant Floquet dressing introduces an additional valley-asymmetric dependence through $\Gamma_\tau$. This immediately introduces a propagation threshold at which the transmission requires a real longitudinal wave vector inside the barrier. Below this threshold, the mode is evanescent, and the current is strongly suppressed. This is the physical explanation of the onset seen in our energy analyses, where transmission activates only when $q_x^{\tau s_z}$ becomes real (Fig.~\ref{fig:T_E}), with the $K$ and $K^{\prime}$ thresholds shifted in opposite directions by the renormalized gap $\Gamma_\tau$.

The same physics explains why electrostatic barriers behave so differently in the two materials. In pure graphene, valley channels are related by time-reversal symmetry and contribute almost identically, and spin splitting is negligible in typical transport conditions \cite{Neto2009, Min2006}. Therefore, an electrostatic barrier mainly produces interference fringes but with weak intrinsic selectivity. To obtain robust filtering in graphene, one usually needs extra symmetry breaking, for example, magnetic barriers, substrate-induced gaps, or proximity-enhanced spin-orbit coupling \cite{Masir2013, Min2006, Mekkaoui2015}. While in monolayer MoS$_2$, spin- and valley are already coupled at the band edge, so scattering phases and resonance conditions depend explicitly on $\tau s_z$ even for a purely electrostatic barrier \cite{xiao2012coupled, Schaibley2016}.
This is also where our approach fits naturally into the existing control strategies for TMDs transport. Spin-valley selectivity can be achieved by magnetic barriers \cite{jellal2025magbarrier}, and by gate-defined junctions where Fabry-P\'erot physics produces strong modulation \cite{cheng2015transport}. Here,  we have added an optical control knob. Off-resonant Floquet dressing renormalizes the effective mass in a {valley-and} polarization dependent way. It shifts both the propagation condition and the Fabry-P\'erot phase, and does so differently for each {valley} channel. This is why a single driven MoS$_2$ barrier can be switched between broadband and resonance-selective filtering by tuning $(\xi, I_L)$.

From a device point of view, graphene excels in ultrahigh-mobility transport but its strong barrier transparency makes robust electrostatic switching difficult \cite{geim2009graphene,Neto2009}. Monolayer MoS$_2$ provides a direct gap and intrinsic spin-valley coupling. It enables gate-defined and optically reconfigurable functionalities relevant to valleytronic and optoelectronic applications \cite{radisavljevic2011single,mak2014valley,sie2015stark}.

\section{Conclusion}\label{S7}

We studied spin- and valley-resolved transport through a single electrostatic barrier in a  monolayer MoS$_2$ with a direct gap $\Delta=1.8$~eV under an off-resonant elliptically polarized irradiation. We used the massive Dirac model with intrinsic spin-orbit coupling and treated the periodic drive within the high-frequency Floquet expansion. This leads to an effective static Hamiltonian where the light field renormalizes the mass (gap) term in a controlled  valley-asymmetric way. This gap renormalization is tuned by two distinct parameters, the laser amplitude $A_0$ and the polarization shape $\xi$.
Solving the barrier-scattering problem allowed us to obtain an exact analytical expression for the transmission probability. 
Our numerical results show that the filtering mechanism is governed by two correlated ingredients.  First, a propagation threshold occurs when the longitudinal wavevector inside the barrier transitions from an evanescent to a propagating regime. Second, in the propagating mode, the multiple reflections within the barrier interfaces generate Fabry-P\'erot resonances, which are set by the phase $q_x^{\tau s_z}L=n\pi$. The applied Floquet dressing modifies both the threshold and the accumulated phase. The key feature here is that this modification 
{is valley-asymmetric, encoded in $\Gamma_\tau$, which increases the gap of the $K$ valley and decreases that of the $K^{\prime}$ valley}. Our results demonstrate that a wide control window emerges where one 
{valley} dominates the transmitted current, while at the same time, the 
{the competing valley is} strongly suppressed. We demonstrated that by tuning $(I_L,\xi)$, the same junction can be reconfigured between broadband filtering (robust pass/stop windows) and resonance-selective operation (narrow, phase-controlled pass bands).

Our findings for the conductance confirm that this selectivity remains visible even after angular averaging, which is the most relevant measure for device operation. In the parameter ranges explored here, the response is mainly valley selective, while the net spin contrast is weaker. {Quantitatively, the valley conductance contrast reaches approximately $18\%$, with $G_K\simeq 1.23$ and $G_{K'}\simeq 0.86$ in the weak-drive limit. The Floquet gap renormalization tunes the propagation thresholds continuously 
	{across the two valleys, with the $K$ and $K^{\prime}$ thresholds shifting in opposite directions}. The broadband-to-resonance-selective transition occurs near the critical amplitude $A_0^{\rm c}\simeq 
	0.55$~V fs/nm, and 
	{a valley-selective optical switch-off is achieved, in which the $K$ conductance collapses to nearly zero while the $K^{\prime}$ conductance remains close to unity,} at $A_0\simeq 0.8$--$0.9$~V fs/nm for circular 
	polarization, which offers a $100 \%$ valley polarization in this regime}
	
	Finally, comparison with graphene highlights the advantage of MoS$_2$, where its intrinsic gap and strong spin-orbit coupling naturally create channel-dependent thresholds and phases. However, graphene typically requires additional symmetry breaking to obtain comparable selectivity knobs. In general, a driven MoS$_2$ barrier provides a compact and optically tunable platform for generating and controlling valley-polarized currents. It offers a practical route toward optically reconfigurable elements for valleytronics and spintronics /optoelectronic device applications.
	
	\section*{Acknowledgment}
	K. Azaidaoui acknowledges the support provided by CNRST in the framework of the program "PhD-Associate Scholarship –- PASS".
	H. Bahlouli acknowledges the support provided by the Interdisciplinary Research Center (IRC) for Advanced Materials and King Fahd University of Petroleum \& Minerals (KFUPM).

	\appendix
	
	\section{Determining transmission}
	\label{appendix}

	We present the detailed derivation of the transmission probability. To determine the transmission probability, we impose the continuity conditions of the eigenspinors, given in ~(\ref{phi1}), (\ref{phi3}), and (\ref{phi2}), across the three regions at $x=0$ and $x=L$.
	These conditions ensure the proper matching of the wave functions at the interfaces and are given by
	\begin{align}\label{eq:BC}
		&\phi_{\text{I}}^{\tau s_z}(0)=
		\phi_{\text{II}}^{\tau s_z}(0),
		\\
		&\phi_{\text{II}}^{\tau s_z}(L)=
		\phi_{\text{III}}^{\tau s_z}(L).
	\end{align}
	They can be explicitly written as
	\begin{widetext}
		\begin{align}
			&\binom{1}{s\alpha^{\tau s_z}e^{i\theta^{\tau s_z}}}+r^{\tau s_z}\binom{1}{-s\alpha^{\tau s_z}e^{-i\theta^{\tau s_z}}}
			=a^{\tau s_z}\binom{1}{s'\beta^{\tau s_z}e^{i\varphi^{\tau s_z}}}+b^{\tau s_z}\binom{1}{-s'\beta^{\tau s_z}e^{-i\varphi^{\tau s_z}}},\\[4pt]
			&a^{\tau s_z}\binom{1}{s'\beta^{\tau s_z}e^{i\varphi^{\tau s_z}}}e^{iq_x^{\tau s_z}L}+b^{\tau s_z}\binom{1}{-s'\beta^{\tau s_z}e^{-i\varphi^{\tau s_z}}}e^{-iq_x^{\tau s_z}L}
			=t^{\tau s_z}\binom{1}{s\alpha^{\tau s_z}e^{i\theta^{\tau s_z}}}e^{ik_x^{\tau s_z}L}.
		\end{align}
	\end{widetext}
	Expanding these equalities, one obtains a set of four linear equations relating the coefficients 
	$r^{\tau s_{z}}$, $a^{\tau s_{z}}$, $b^{\tau s_{z}}$, and $t^{\tau s_{z}}$ as
	\begin{widetext}
		\begin{align}
			&1+r^{\tau s_z}=a^{\tau s_z}+b^{\tau s_z},
			\label{eq:bc1}\\
			&s\alpha^{\tau s_z}e^{i\theta^{\tau s_z}}
			-s\alpha^{\tau s_z}r^{\tau s_z}e^{-i\theta^{\tau s_z}}
			=a^{\tau s_z}s'\beta^{\tau s_z}e^{i\varphi^{\tau s_z}}
			-s'\beta^{\tau s_z}b^{\tau s_z}
			e^{-i\varphi^{\tau s_z}},
			\label{eq:bc2}\\
			&a^{\tau s_z}e^{iq_x^{\tau s_z}L}
			+b^{\tau s_z}e^{-iq_x^{\tau s_z}L}
			=t^{\tau s_z}e^{ik_x^{\tau s_z}L},
			\label{eq:bc3}\\
			&a^{\tau s_z}s'\beta^{\tau s_z}
			e^{i\varphi^{\tau s_z}}e^{iq_x^{\tau s_z}L}
			-b^{\tau s_z}s'\beta^{\tau s_z}
			e^{-i\varphi^{\tau s_z}}e^{-iq_x^{\tau s_z}L}
			=t^{\tau s_z}s\alpha^{\tau s_z}
			e^{i\theta^{\tau s_z}}e^{ik_x^{\tau s_z}L}.
			\label{eq:bc4}
		\end{align}
	\end{widetext}
	In matrix form, we have
	\begin{widetext}
		\begin{equation}
			\begin{pmatrix}
				-1 & 1 & 1 & 0\\[2pt]
				s\alpha^{\tau s_z}e^{-i\theta^{\tau s_z}} & s'\beta^{\tau s_z}e^{i\varphi^{\tau s_z}} & -s'\beta^{\tau s_z}e^{-i\varphi^{\tau s_z}} & 0\\[2pt]
				0 & e^{iq_x^{\tau s_z}L} & e^{-iq_x^{\tau s_z}L} & -e^{ik_x^{\tau s_z}L}\\[2pt]
				0 & s'\beta^{\tau s_z}e^{i\varphi^{\tau s_z}}e^{iq_x^{\tau s_z}L} & -s'\beta^{\tau s_z}e^{-i\varphi^{\tau s_z}}e^{-iq_x^{\tau s_z}L} & -s\alpha^{\tau s_z}e^{i\theta^{\tau s_z}}e^{ik_x^{\tau s_z}L}
			\end{pmatrix}
			\begin{pmatrix}
				r^{\tau s_z}\\[2pt] a^{\tau s_z}\\[2pt] b^{\tau s_z}\\[2pt] t^{\tau s_z}
			\end{pmatrix}
			=
			\begin{pmatrix}
				1\\[2pt] s\alpha^{\tau s_z}e^{i\theta^{\tau s_z}}\\[2pt] 0\\[2pt] 0
			\end{pmatrix}
		\end{equation}
		Inverting the above matrix yields 
		the transmission and reflection coefficients 
		\begin{align}
			t^{\tau s_z}
			&= \frac{s s' e^{i k_x^{\tau s_z} L} \cos\theta^{\tau s_z} \cos\varphi^{\tau s_z}}
			{s s'\cos\big(q_x^{\tau s_z}L\big)\cos\theta^{\tau s_z}\cos\varphi^{\tau s_z}
				-i\sin\big(q_x^{\tau s_z}L\big)\left(\frac{(\alpha^{\tau s_z})^2+(\beta^{\tau s_z})^2}{2\alpha^{\tau s_z}\beta^{\tau s_z}}
				- s s'\sin\theta^{\tau s_z}\sin\varphi^{\tau s_z}\right)},\label{ttsz}\\
			r^{\tau s_z}
			&= \frac{\sin\big(q_x^{\tau s_z}L\big) e^{i\theta^{\tau s_z}}
				\left[s s'\sin\varphi^{\tau s_z}
				-i\frac{(\alpha^{\tau s_z})^2 e^{i\theta^{\tau s_z}}-(\beta^{\tau s_z})^2 e^{-i\theta^{\tau s_z}}}
				{2\alpha^{\tau s_z}\beta^{\tau s_z}}\right]}
			{s s'\cos\big(q_x^{\tau s_z}L\big)\cos\theta^{\tau s_z}\cos\varphi^{\tau s_z}
				-i\sin\big(q_x^{\tau s_z}L\big)\left(\frac{(\alpha^{\tau s_z})^2+(\beta^{\tau s_z})^2}{2\alpha^{\tau s_z}\beta^{\tau s_z}}
				- s s'\sin\theta^{\tau s_z}\sin\varphi^{\tau s_z}\right)}.
		\end{align}
	\end{widetext}
	The transmission probability follows from the expectation value of the current operator, which in the $x$-direction is given by
	\begin{equation}
		J_x^{\tau s_z}=v_F\phi^{\tau s_z\dagger}\tau\sigma_x\phi^{\tau s_z} ,\label{ccdd}
	\end{equation}
	By substituting the spinor solutions (\ref{phi1}), (\ref{phi2}), and (\ref{phi3}) into \eqref{ccdd}, we show that  the incident, reflected, and transmitted currents take the forms
	\begin{align}
		J_{\mathrm{in}}^{\tau s_z}&=2s\tau v_F\cos\theta^{\tau s_z},\\
		J_{\mathrm{ref}}^{\tau s_z}&=-2s\tau v_F|r^{\tau s_z}|^2\cos\theta^{\tau s_z},\\
		J_{\mathrm{tr}}^{\tau s_z}&=2s\tau v_F|t^{\tau s_z}|^2\cos\theta^{\tau s_z}.
	\end{align}
	Using the ratio $T^{\tau s_z}=|J_{\mathrm{tr}}^{\tau s_z}|/|J_{\mathrm{in}}^{\tau s_z}|=|t^{\tau s_z}|^2$, we obtain the explicit expressions for the  transmission probabilities
	\begin{widetext}
		\begin{align}
			T^{\tau s_z} &= \frac{ \cos^2 \theta^{\tau s_z}\cos^2 \varphi^{\tau s_z}}{ \cos^2\left(q_x^{\tau s_z}L \right) \cos^2  \theta^{\tau s_z} \cos^2 \varphi^{\tau s_z}+\sin^2\left(q_x^{\tau s_z}L \right)\left(\frac{(\alpha^{\tau s_z})^2+(\beta^{\tau s_z})^2}{2\alpha^{\tau s_z}\beta^{\tau s_z}}-ss^{\prime} \sin\theta^{\tau s_z} \sin\varphi^{\tau s_z}\right)^2}.
		\end{align}
	\end{widetext}
	
\end{document}